\documentclass[aps,prb,twocolumn,a4paper,eqsecnum,balancelastpage,showpacs,floatfix]{revtex4}
\usepackage{epsfig}
\def\dir{}

\begin{document}
\title{Role of surface waves on the relation between crack speed and the work of fracture}
\author{Andrea Parisi}
\author{Robin C. Ball}
\affiliation{Department of Physics, University of Warwick, Coventry 
CV4 7AL, United Kingdom}

\date{\today}
\begin{abstract} 
We show that the delivery of fracture work to the tip of an advancing planar 
crack is strongly reduced by surface phonon emission, leading to forbidden 
ranges of crack speed.  The emission can be interpreted through dispersion 
of the group velocity, and Rayleigh and Love branches contribute as well as 
other high frequency branches of the surface wave dispersion relations.
We also show that the energy release rate which enters the Griffith criterion
for the crack advance can be described as the product of the continuum solution
with a function that only depends on the lattice geometry and describes
the lattice influence on the phonon emission.
Simulations are performed using a new finite element model for simulating 
elasticity and fractures.  The model, built to allow fast and very large 
three-dimensional simulations, is applied to the simplified case of two 
dimensional samples.
\end{abstract}


\maketitle

%
%

\section{INTRODUCTION}

It is well known that fracture propagation is strongly governed by properties of 
the surface created.  
The classical Griffith criterion requires that for a crack to propagate all the
elastic energy supplied to the crack tip must at least match the work of surface
creation.  Crack speed is more complicated: one needs to understand the
energetics of mechanisms in competition with surface work, and in this paper we
focus on phonon emission.

Phonon emission from the crack tip involves both bulk and surface waves.
In the continuum limit these waves are dispersionless and emission is only 
expected when the crack speed matches any of the speeds $v_l$, 
$v_t$ and $v_R$ of longitudinal, transverse and Rayleigh (surface) waves.  The
Rayleigh speed being the lowest, it is often regarded as the ultimate theoretical
limit for the speed of crack propagation.
Real materials however have properties that do not match the continuum theory.  
One famous example of this is the existence of forbidden crack speeds, which 
was first explained by M.~Marder \emph{et al.}\cite{MG-95,HM-99,FM-99} taking 
into account the discrete nature of matter.
Their qualitative explanation of the \emph{velocity gap} where no crack 
can propagate in a periodic lattice at speeds lower than roughly $1/3$ of 
the transverse waves speed, 
was related to lattice 
oscillations breaking bonds before the expected time compatible with the 
crack speed.  
This explanation involves the use of a ``most stretched bond'' breakage rule.  
Here we propose a more cautious but general view using an approach based on 
the phonon band structure and the energy release rate.  
The phonon band structure can lead to resonant emissions which influence the 
fraction of energy radiated from the crack tip.  The existence of forbidden 
crack speeds then follows.

When dealing with dispersion relations the reference is usually to bulk waves 
or, in the case of surface waves, to the Rayleigh branch responsible for the 
Rayleigh speed.  
Geophysicists and researchers in earthquakes are also familiar with 
another kind of surface waves known as Love waves\cite{B-54,P-62}, polarized in the direction 
normal to that of propagation and parallel to the crack plane.
Love waves only exist in the continuum limit when there is a gradation of elastic 
properties near the surface, pertinent to seismology but not usually considered in 
fracture mechanics.
However, due to the discrete nature of matter at the atomic level 
dispersive media can also support Love and other wave branches so that they can 
enter fracture problems.
It is natural to include the whole complexity of the surface dispersion 
relations in the description of the material to understand its role, and how and 
if this complexity can change the theoretical description of fracture dynamics.

The form of dispersion relations is the fingerprint of the discrete spatial 
organization of matter, beyond the continuum limit.  In any kind of simulation 
it is necessary to deal with discretization 
in space.  In molecular dynamic simulations\cite{ABRR-94,ZBLH-97,HM-98} this corresponds to 
an atomistic description; in mesoscopic models as 
lattice-like\cite{PGLGS-98,PGLGS-00,MERZ-00,CCG-98} or finite element 
models\cite{CP-89,J-92,XN-94}, 
discretization has to be imposed in space in order to solve differential 
equations of the dynamics.  This has an unavoidable effect on the dispersion 
relations that necessarily reflect the underlying lattice.

In this paper we present a new model developed to simulate an elastic continuum 
material using finite tetrahedral elements.  Space is discretized on an fcc 
grid, and this is reflected in the dispersion relations of the simulated 
material.  
This feature could be considered a drawback, but can be 
exploited to model the influence of similar dispersions in real fcc structure 
materials.  
The simplicity of the model allows us an extensive understanding 
of both simulations and theory.
The model, built to allow fast and very large three dimensional simulations, is 
used here in the simplified case of two dimensional samples, and phonon emission from 
the crack surface of planar cracks advancing at fixed speeds is analysed.  
An analysis of the influence of this emission on the crack dynamics is the aim of this work.  

New and central to our approach is that we fix the crack properties 
(speed and shape) and measure some of the mechanical properties (stress, 
energy release rate).  This can be done with success in two-dimensions and 
gives new insight into the behaviour of some of these quantities.
In particular, we show that the energy release rate is strongly influenced
by the surface phonon emission at the crack tip through 
resonant emission at particular crack speeds, leading to the existence of
narrow bands of permitted crack speed.  We also show that the energy
release rate which enters the Griffith criterion can be expressed as the 
continuum solution, multiplied by a microscopic function that only depends 
on the lattice geometry and that describes the lattice influence on the phonon emission.

The paper is organized as follows:
in section II a full description of the model is provided, including the 
description of the bulk dispersion relations.
In section III, a preliminary analysis of simulations of planar cracks at 
constant speed is given, and in section IV we relate the phonon emission to 
the dispersion relations of surface waves and we show the phenomenon of resonances.  
In section V we analyse how the energy release rate is influenced by the
phonon emission and we derive the existence of bands of permitted crack speeds.
Finally, conclusions and comparison with other works are given in 
section VI.

%
%

\section{SIMULATION MODEL}

The model is an application of the finite element method with the aim of 
simulating linear elasticity and fractures.
In continuum linear elasticity \cite{LL-59} the Lagrangian has the form:
\begin{equation}
\label{eq:lagrangian-cont}
L = \int \bigg( \frac{\rho \dot{\bf u}^2}{2} - \frac{1}{2} \nabla {\bf u} : 
    \stackrel{\leftrightarrow}{\sigma} \bigg) \ dV
\end{equation}

\noindent
where the stress tensor is given by \cite{F-90}:
\begin{equation}
\label{eq:stress-strain}
\stackrel{\leftrightarrow}{\sigma} = \lambda \textrm{Tr(} \nabla {\bf u}
         \textrm{)}\,\openone + \mu [ \nabla {\bf  u} + (\nabla {\bf u})^T ]
.
\end{equation}

\noindent
The unsymmetrized strain tensor $\nabla {\bf u}$ from which the stress tensor,
Lagrangian and hence equations of motion follow is central to our finite element
scheme below.

The key driving features of the model are that the elastic response is as local
as possible so that non-linearity and particularly rupture can be incorporated,
whilst accidental soft modes are avoided and all mode frequencies are strictly 
upper bounded for stability of timestepping.

\subsection{Definition of the elastodynamic model}

\begin{figure}[t]
\epsfig{figure=\dir 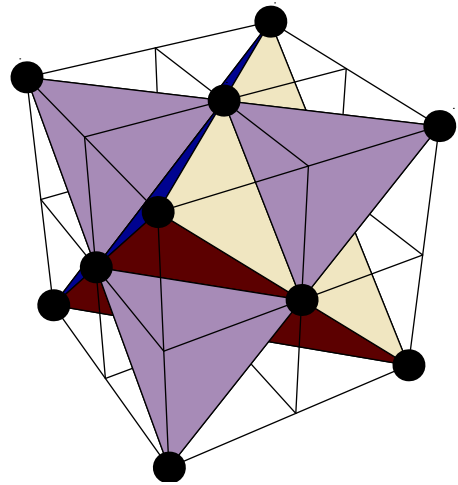, height=6.5cm}
\medskip
\caption{Unit cell for the fcc lattice used in our simulations.  This contains
eight tetrahedra, each connecting four lattice sites.  
Displacement and momentum are defined on the sites, whilst strain
and stress are defined on the tetrahedra.}
\label{fig:FCClattice-tet}
\end{figure}

To evaluate the unsymmetrized strain tensor, 
space is discretized using an fcc lattice and lattice points are
connected using tetrahedral elements.  Each element connects
four points (see fig.~\ref{fig:FCClattice-tet}), which is the minimum 
for a full local gradient calculation in three dimensions.
On each lattice point, the displacement field ${\bf u}$ is defined, and the 
Lagrangian in its discretized form is given by:
\begin{equation}
\label{eq:lagrangian-discr}
L = \sum_v \frac{m \dot{\bf u}_v^2}{2} - \sum_t \frac{1}{2} \Omega' (\nabla 
    {\bf u})_t : \stackrel{\leftrightarrow}{\sigma}_t 
\end{equation}

\noindent
where the $v$ index spans over all lattice points and the $t$ index spans all 
the tetrahedral elements.  $\Omega'$ is the volume of the system per 
tetrahedron, $1/8$ of the volume of the fcc conventional unit cell shown in 
fig.~\ref{fig:FCClattice-tet}.

The stress tensor is evaluated at the center of each tetrahedron through 
(\ref{eq:stress-strain}), given the
unsymmetrized strain tensor at the center of the $t$-th tetrahedron is:
\begin{equation}
\label{eq:strain}
(\nabla {\bf u})_t = \alpha \sum_{v} {\bf r}_{tv} {\bf u}_v
\end{equation}

\noindent
with the condition $\alpha \sum_v {\bf r}_{tv} {\bf r}_{tv} = \openone$,
where ${\bf r}_{tv}$ is the vector joining the center of the $t$-th tetrahedron 
with neighbouring vertex site $v$, and ${\bf u}_v$ is the displacement field on 
site $v$.
Finally, the equation of motion takes then the form:
\begin{equation}
\label{eq:elastic-motion}
m \ddot{\bf u}_v = {\bf f}_v = - \alpha \Omega' \sum_t {\bf r}_{tv} \cdot 
  \stackrel{\leftrightarrow}{\sigma}_t
\end{equation}

\noindent
where $\stackrel{\leftrightarrow}{\sigma}_t$ is
the stress tensor of the tetrahedron.

\begin{figure*}[t!]
\centerline{
\epsfig{figure=\dir 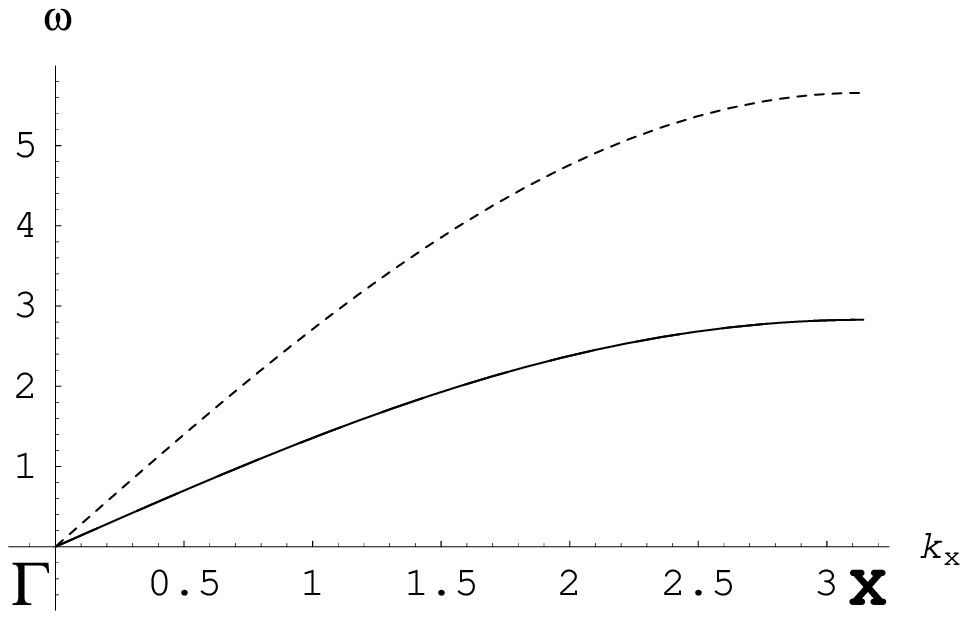, height=4.6cm}
\hspace{1cm}
\epsfig{figure=\dir 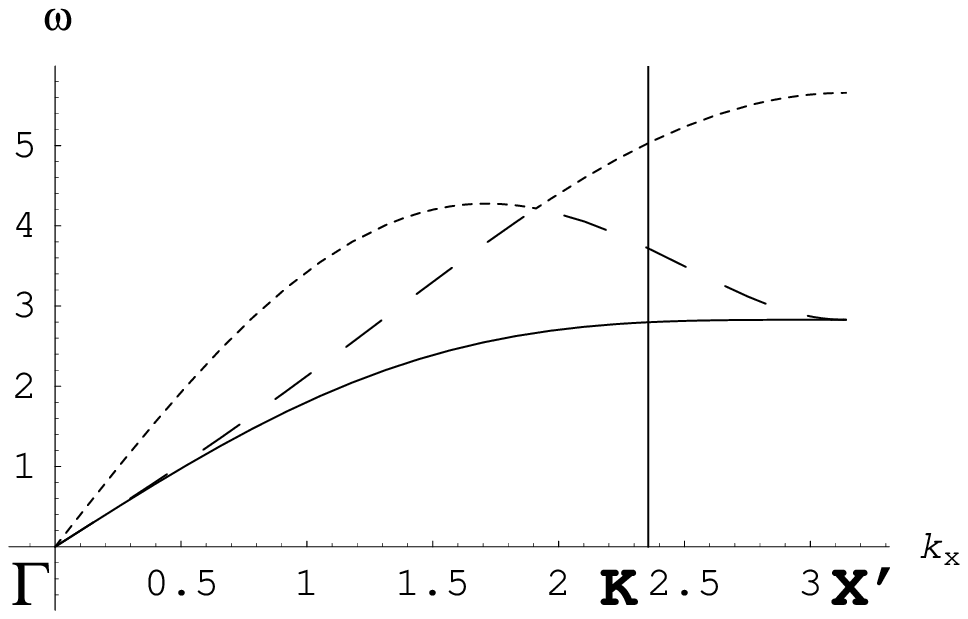, height=4.6cm}}

\centerline{
\epsfig{figure=\dir 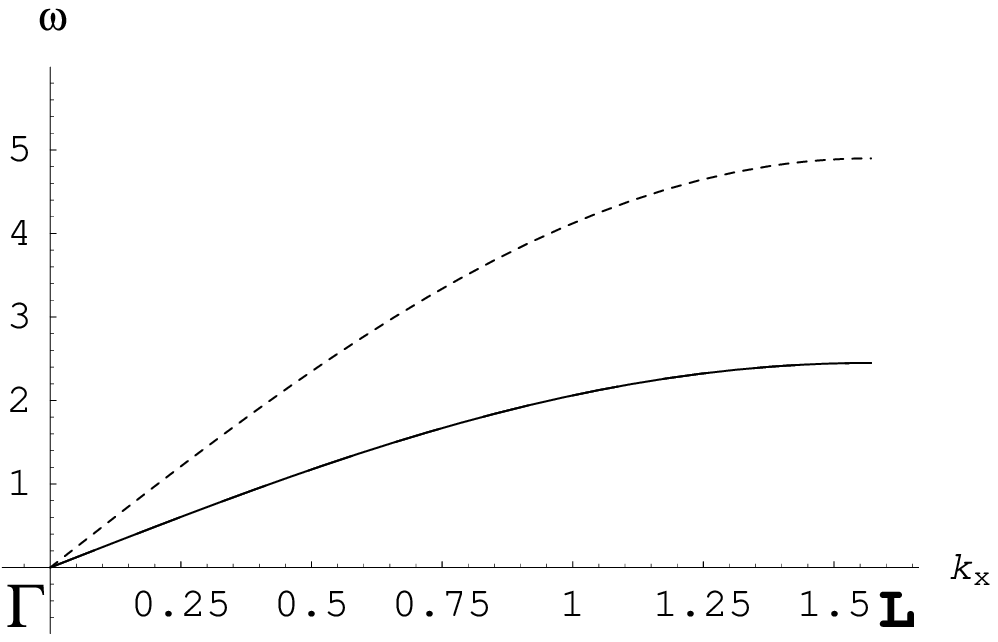, height=4.6cm}
\hspace{1.2cm}
\epsfig{figure=\dir 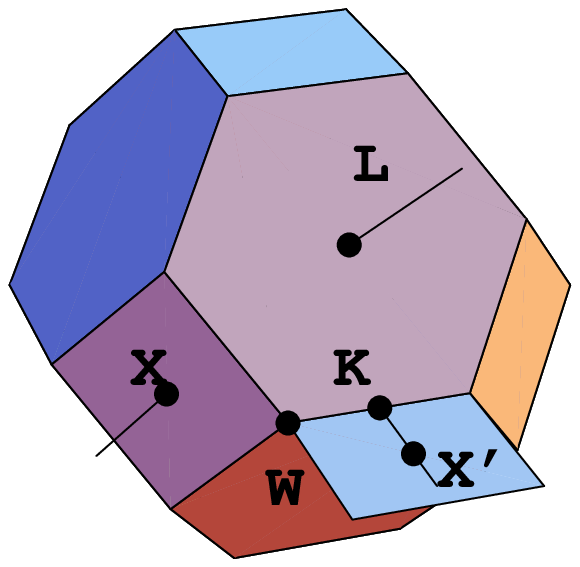, 
height=6.3cm}\hspace{1.1cm}}
\medskip
\protect 
\caption{Acoustic dispersion curves for the tetrahedron model along different 
directions in reciprocal space: there are no soft modes at the border of the 
Brillouin Zone.  These plots are shown for Poisson's ratio $\nu=1/3$ and using
units where $a=1$ for the side of the tetrahedral cell and $v_t=\protect\sqrt{2}$ for
the speed of transverse acoustic waves.}
\label{fig:FCClattice-disp-rel}
\end{figure*}

%
%

\subsection{Dispersion relations for bulk waves}

The dispersion relations of our model are relatively straightforward to
calculate.  These confirm that in the long wavelength limit 
the continuum case is retrieved, and particularly that the use of tetrahedral elements avoids the
presence of soft modes at the border of the Brillouin zone, a problem arising when using simple
cubic schemes.

Introducing a spatial Fourier transform in equation (\ref{eq:strain}), the 
gradient operator can be written as:
\begin{equation}
\label{eq:FT-nabla}
\textrm{FT}[\nabla] = \alpha \sum_{ {\bf r}_v } {\bf r}_v
                      e^{i {\bf k} \cdot {\bf r}_v}
.
\end{equation}

There are two kinds of tetrahedra (see fig.~\ref{fig:FCClattice-tet}) which can
be readily seen to be mirror images of each other, and their corresponding
gradient operators are related, in the Fourier transform, by 
$\nabla_1 = - \nabla_2^*$.  In terms of these eq.~(\ref{eq:elastic-motion}) 
can be expressed as:
\begin{equation}
\label{eq:rho-sigma}
m \ddot{\bf u} = \Omega' \Big[ \nabla_2 \cdot 
\stackrel{\leftrightarrow}{\sigma}_1 +
                    \nabla_1 \cdot \stackrel{\leftrightarrow}{\sigma}_2 \Big]
\end{equation}

\noindent
where
\begin{equation}
\label{eq:sigma-1}
\stackrel{\leftrightarrow}{\sigma}_1 =
          \lambda \stackrel{\leftrightarrow}{I} \nabla_1 \cdot {\bf u} +
          \mu \Big[\nabla_1 {\bf u} + {\bf u} \nabla_1 \Big]
\end{equation}

\noindent
and $\stackrel{\leftrightarrow}{\sigma}_2$ is obtained by changing the index
from $1$ to $2$ in eq.~(\ref{eq:sigma-1}).

Equation (\ref{eq:rho-sigma}) is therefore given, after Fourier
transforming, by
\begin{eqnarray}
\label{eq:motion-model}
\lefteqn{m \omega^2 {\bf u}_{\bf k} = \Omega' \Bigg\{ \mu \Big[\nabla \cdot
         \nabla^* + \nabla^* \cdot \nabla \Big] {\bf u}_{\bf k} + }
	 \hspace{1.5cm}\nonumber\\
 &  & + (\lambda + \mu) \Big[\nabla \nabla^* + \nabla^* \nabla \Big] \cdot 
         {\bf u}_{\bf k} \Bigg\}
\end{eqnarray}

\noindent
where $\nabla = \nabla_1 = - \nabla_2^*$.
This equation can be solved exactly for eigenvalues (see appendix
\ref{app:bulk-waves}), giving the results shown graphically in figure
\ref{fig:FCClattice-disp-rel}.
The dispersion curves are well
behaved, and reflect the underlying fcc structure: even if we are trying to 
simulate the continuum limit, dispersion relations are the image of the spatial 
structure of the underlying lattice.  This cannot be avoided in our model just 
as in real materials.

The dispersion relations found vindicate the use of tetrahedral elements.
Using a simpler cubic lattice and cubic finite elements leads to unphysical 
soft modes, where the frequency goes to zero whenever
two of the three components of the ${\bf k}$ vector approach the border of the 
Brillouin zone.
The use of tetrahedra
removes this problem since there is no way to deform a tetrahedron giving a null 
contribution to the strain tensor, whilst alternative strategies involving 
less local spatial derivatives make fracture properties harder to implement 
and control.

The value of the maximum frequency is set by the longitudinal branch at the 
Brillouin Zone boundary.  
We will see in section \ref{sec:maxfreq} that such a value remains an
upper bound even when considering the finiteness of the sample or the presence 
of fractures.  This value provides the minimum period of site motion.

Due to the significant role of surface waves in this paper, a study on their dispersion 
relations will be described later in section \ref{sec:surf-disp-rel}.

%
%

\subsection{Discretizing time}

Time can be discretized by introducing a finite timestep $h$.  The integration
scheme used in our simulations is the \emph{Leap-Frog} scheme
\cite{AT-87}, in which displacements ${\bf u}_{v}$ and momenta ${\bf p}_{v}$
are evaluated alternately at subsequent steps:
\begin{displaymath}
\left\{
\begin{array}{lll}
{\bf u}_{v} (t)  & = & {\bf u}_{v}(t-2h) + 
                           2h\displaystyle\frac{{\bf p}_{v}(t-h)}{m}\\
                   &   & \\
{\bf p}_{v}(t+h)  & = & {\bf p}_{v}(t-h) + 
                           2h\displaystyle{\bf f}_{v}(t)
.
\end{array}
\right.
\end{displaymath}

Fourier transforming these equations and comparing with the limit $h\rightarrow 0$
leads to the dispersion relation:
\begin{equation}
\label{eq:disp-h}
\sin^2(\omega h) = h^2 \omega_0^2({\bf k})
\end{equation}

\noindent
where $\omega_0({\bf k})$ is the value of the frequency of mode ${\bf k}$ 
in the continuous time ($h\rightarrow 0$) limit.
From this equation it follows that we require $h \omega_0(\bf{k}) < 1$
for mode k to be stable in the discrete timestep simulations, and the 
global stability limit is
\begin{displaymath}
h < \Big[ \max_{\bf{k}}\, \omega_0(\bf{k}) \Big]^{-1}
.
\end{displaymath}

It is interesting to apply this to the simplified case of one dimension. 
In this case the continuous time dispersion relations have the simple form
\begin{equation}
\label{eq:disp-1d}
\omega_0^2(k) = \omega^2_{\textrm{\small max}} \sin^2 \Big(\frac{ka}{2}\Big)
\end{equation}

\noindent
with: $\omega_{\textrm{\small max}} = (2/a)
\sqrt{(\lambda + 2\mu)/{\rho}}$.
Comparing (\ref{eq:disp-h}) and (\ref{eq:disp-1d}), it is easy to observe 
that the stability limit for the timestep is $h=1/\omega_{\textrm{\small max}}$.
Moreover, if we use exactly this value for the timestep, we obtain the 
linear dispersion relation $\omega = ck$ with
$c = a\,\omega_{\textrm{\small max}}/{2}$.
Thus using the maximum allowed timestep, the relation between the
frequency and the wave vector is the same as in continuum elasticity.
This notable property is valid only in one dimension and is lost
in two and three dimensions where the dispersion is more rich.
Nevertheless, this property shows that having 
discretized space, it not necessarily an optimal approximation to the continuum 
to use a very small timestep as in standard Molecular Dynamics simulations.

In three dimensions $\omega_{\textrm{\small max}}$ is given by the maximum
frequency over the entire first Brillouin zone.
Using the maximum allowed timestep dispersion relations are closer but not
equivalent to the continuum limit.  

The Leap-Frog scheme used here to discretize time is relatively robust with
respect to energy conservation due to its time reveribility and symplectic 
properties.\cite{FS-02}  There is an exactly conserved $h$-dependent hamiltonian $H_h$
very close to the na\"{\i}ve one\cite{FS-02}, with for example:
\begin{equation}
H_h \equiv \frac{1}{2} \Big( K_+ + K_- \Big) + U + \textrm{O}( h^2 )
\end{equation}

\noindent
where $K_+$ and $K_-$ are the kinetic energies advanced and retarded by $\pm h$
relative to the timestep where the potential energy $U$
is evaluated.

In these simulations the value chosen for the timestep has been 
$h = 0.1\ h_{\textrm{\small max}}$.  Using such timestep, dispersion relation 
are close to the continnum time dispersion relations.
Some trial simulations with $h = 0.01\ h_{\textrm{\small max}}$ have been 
performed showing no substantial difference with the results here reported.

%
%

\subsection{Nonlinear elasticity and breakage}
\label{sec:maxfreq}
The model up to now has been described as a model for linear elasticity.
However, since the link between the stress and the strain tensor is given
exclusively through equation (\ref{eq:stress-strain}), we can easily generalize
the equation to include any kind of elastic response as well as anisotropies.
For instance:
\begin{equation}
\stackrel{\leftrightarrow}{\bf \sigma} =
    \stackrel{\leftrightarrow}{\bf f}\!\!\Big(\nabla {\bf u} + 
             (\nabla{\bf u})^T \Big)
\end{equation}

\noindent
where each tensor component $f_{ij}$ represents any kind of nonlinear function.
In this paper we study only the simpler scenario of linear elasticity augmented
by breakage of tetrahedra corresponding to the advance of the crack.  
We brake tetrahedra by abruptly setting its elastic constants $\lambda$ and
$\mu$ to zero and it is the resulting recoil of the neighbouring sites which
excites phonon emission.

Breaking tetrahedra does not compromise the maximum allowed timestep, because we
can readily show that the maximum vibrational frequency cannot increase.  To see
this consider eq.~(\ref{eq:elastic-motion}) rewritten in the form

\begin{displaymath}
\rho \ddot{\bf u}_v = - \frac{\partial {\cal U}}{\partial {\bf u}_v}
.
\end{displaymath}

\noindent
From (\ref{eq:lagrangian-discr}), we clearly have 
$\sum_v {\bf u}_v \cdot \partial {\cal U} / \partial {\bf u}_v = 2 {\cal U}$, 
where ${\cal U}$ is the total potential energy.  
Now consider a normal mode obeying:
\begin{displaymath}
\rho \omega^2 {\bf u}_v = \displaystyle
     \frac{\partial {\cal U}}{\partial {\bf u}_v}
\end{displaymath}

\noindent
from which it follows that
\begin{displaymath}
\rho \omega^2 = \displaystyle 
     \frac{2 {\cal U}}{\sum_{v} |{\bf u}_v|^2}
.
\end{displaymath}

\noindent
Breaking a tetrahedron removes strictly non-negative terms from the numerator of
the right hand side, and this can only lower the maximum of $\omega$.

%
%

\section{SIMULATIONS OF PLANAR CRACKS AT CONSTANT SPEED}

An advancing crack is simulated through the breaking of tetrahedra.  
To decide when to break we need a \emph{breakage rule}.
One possibility is to break tetrahedra as soon as they overcome some
fracture criterion, for example a critical value of principal stress or elastic energy.
In this work we instead follow a complementary approach, advancing a straight crack at
constant speed and measuring the fracture criteria achieved such as work of fracture.
This provides new insight into the relationship between energy release rate and crack speed, 
including any intervals of forbidden velocities.

\begin{figure}[t]
\centerline{
\epsfxsize=5cm{\epsfbox{\dir 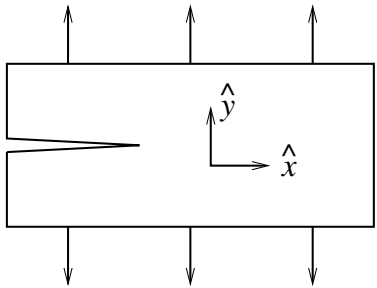}}}
\medskip
\caption{Set-up for simulations.  A fixed displacement is applied on the top and 
bottom boundaries, and the crack advances at constant speed up to the middle of 
the sample.
}
\label{fig:Planar-setup} 
\end{figure}

For computational convenience we focussed on crack speeds commensurate with our
timestep, that is

\begin{displaymath}
v_n= \displaystyle\frac{1}{2hn}
\end{displaymath}

\noindent
for $n=1, 2, 3, \ldots\ $.
Any desired crack speed can be obtained from this sequence by modest 
adjustment of the timestep $h$.

%
%

\subsection{Stress field of advancing cracks}

The model presented above is fully three dimensional.
However for a strictly planar type-I crack, commensurate with our lattice,
it is readily shown that the resulting dynamical solution is strictly planar
with the displacement all in that plane.
We have exploited this and shrunk our system to one fcc cell deep in the
third dimension, with periodic boundary conditions.

In our simulations the Lam\`e coefficients are set so that the Poisson ratio is $\nu=1/3$.
The boundary conditions correspond to imposing fixed normal counter displacement 
at the top and bottom boundaries, i.~e.~$u_x=0$, $u_y=\pm a$ at $y=\pm L/2$, 
and $\sigma_{xx}=\sigma_{xy}=0$ on the right and left faces 
(see fig.~\ref{fig:Planar-setup}). 
The initial condition is the static elastic solution, found by 
relaxing the lattice to its configuration of minimum energy by 
adding dissipation as discussed
in appendix \ref{app:damp}. 
Because of the linearity of the equations, the magnitude of the 
imposed boundary displacement does not influence the results: 
for the following results the starting displacement is $1\,\%$ 
of the sample height. 
The advancing fracture is simulated until it reaches the middle 
of the sample, then a snapshot of the stress field is taken. 

\begin{figure*}[t!]
\centerline{
\epsfysize=6.5cm{\epsfbox{\dir 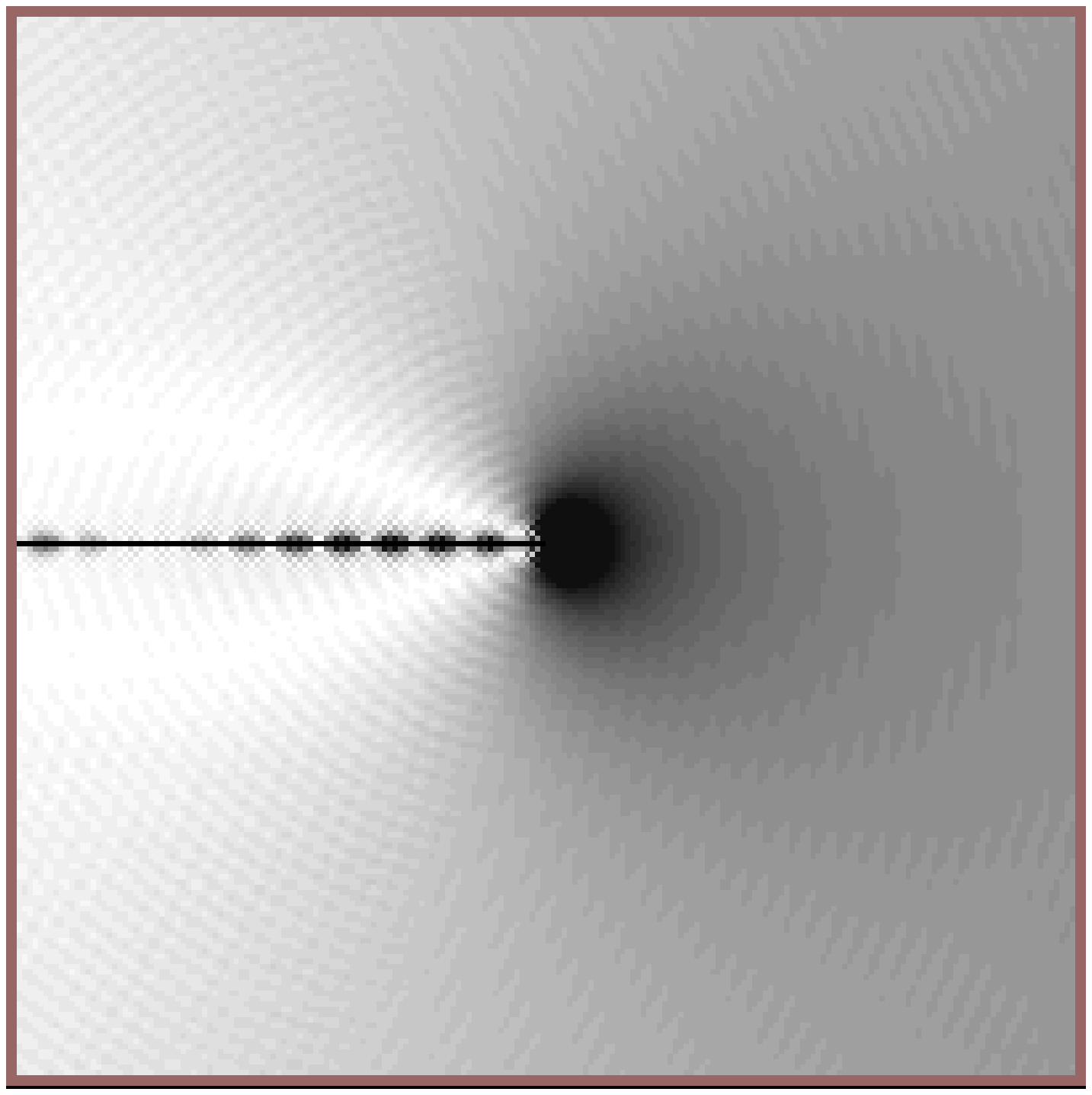}}
\hspace{1cm}
\epsfysize=7cm{\epsfbox{\dir 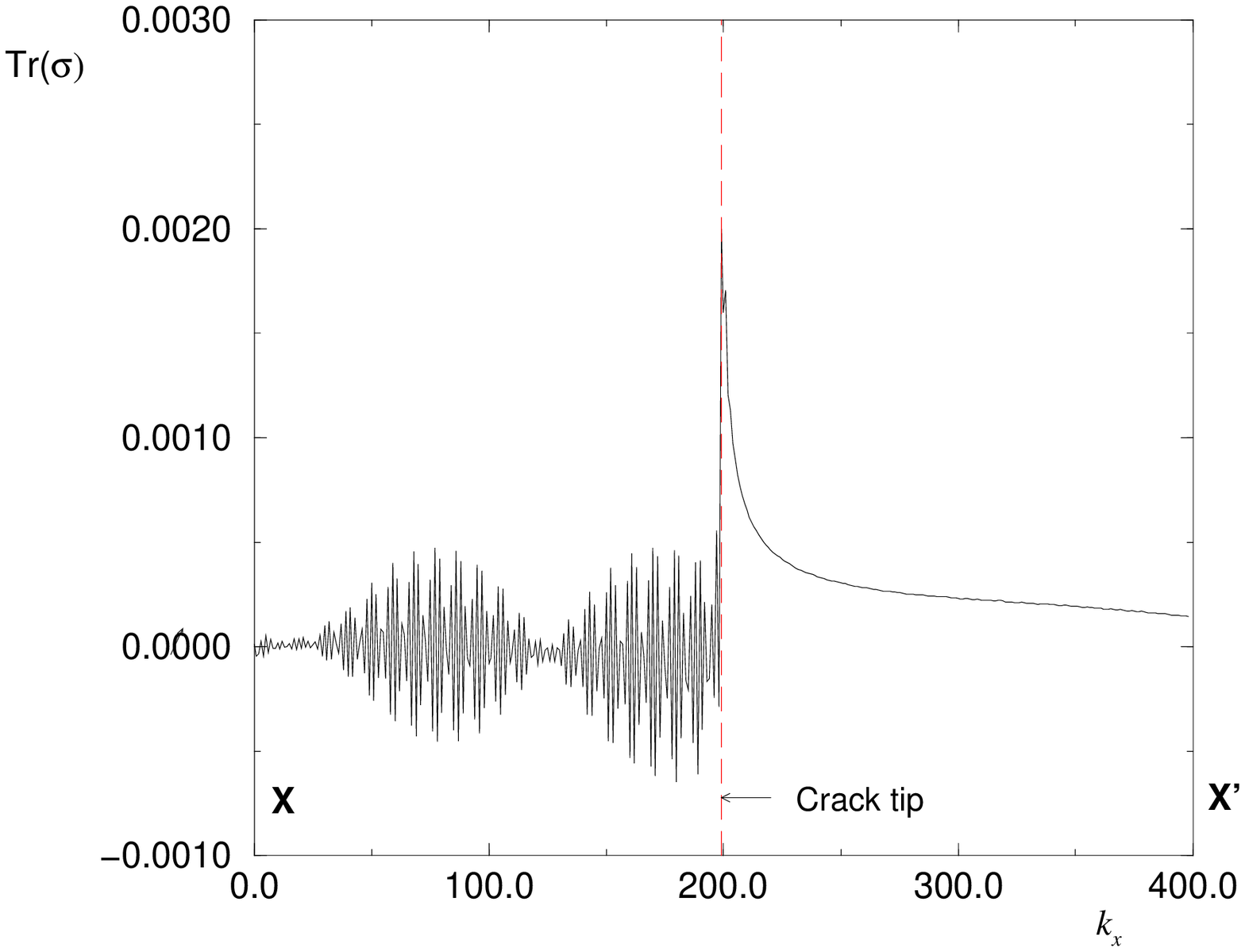}}}
\medskip
\caption{(a) Absolute value of the trace of the stress field for a crack 
advancing at half of the 
transverse wave speed, with Poisson ratio $\nu = 1/3$.  This is the central
$200 \times 200$ region from a sample of $400\times 400$ tetrahedra.  
Intensity spans from white (zero)
to black (highest values).  (b) Trace of the stress field 
measured along the fracture plane, through the crack tip whose singularity 
is clearly visible, and onwards ahead of the crack tip.
}
\label{fig:trace-stressfield} 
\end{figure*}

In figure \ref{fig:trace-stressfield}.(a) the trace of the stress field for a 
typical simulation is shown: 
strong emission of surface waves is visible on the crack surface, 
as also shown in figure \ref{fig:trace-stressfield}.(b).  In particular, from 
this last figure it can be seen that the amplitude of these oscillations is 
smaller but of the same order of magnitude as the stress level at the crack 
tip.

%
%

\subsection{Description of the basic phenomenon}

When a crack is advancing new surface is being created, and waves are emitted
both into the bulk and along the crack surface.
In a system of reference comoving with the crack tip, the frequency $\omega'$ of
emitted waves has to match the temporal frequencies with which lattice structure 
presents itself to the crack tip, leading to the selection rule:
\begin{displaymath}
\omega' = \omega - {\bf k} \cdot {\bf v} = {\bf g} \cdot {\bf v}
\end{displaymath}

\noindent
where ${\bf g}$ is any reciprocal lattice vector.
This is equivalent to a simple matching of phase velocity:
\begin{equation}
\label{eq:cherenkov}
\omega = {\bf k} \cdot {\bf v}
\end{equation}

\noindent
where the wave vector ${\bf k}$ is viewed in the extended zone scheme as 
sketched in fig.~\ref{fig:surf-phenom}.

From a graphical point of view, a crack advancing at a given speed
can be represented as a straight line in the $(k_x,\omega)$ plane,
its slope given by the crack speed.
The crack emits waves at the frequencies and wavelengths corresponding
to the intercepts of this line with the dispersion relations
(see fig.~\ref{fig:surf-phenom}) in the extended zone scheme.

For some crack speeds condition
(\ref{eq:cherenkov}) may be fulfilled not just for one 
${\bf k}$ point, but over a neighbourhood of ${\bf k}$.
This can readily be shown to reduce to:
\begin{equation}
\label{eq:resonant-condition}
\displaystyle {\bf v}_G = \frac{\partial \omega}{\partial {\bf k}} = {\bf v}
.
\end{equation}

\noindent
This matching of phase velocity and group velocity (sketched as well in 
fig.~\ref{fig:surf-phenom}) is the condition of resonant emission.  We can 
expect for these crack speeds a sharp increase in the intensity of the 
phonon emission.

To take this discussion further we need to describe the dispersion relations 
of surface waves for our model, as detailed in the following section.

%
%

\section{DISPERSION RELATIONS FOR SURFACE WAVES}
\label{sec:surf-disp-rel}

The description of the modes in which a lattice with an underlying fcc symmetry 
can vibrate is more complex than the one given by continuum elasticity.  
Frequency as a function of wave vector is no longer linear, and becomes periodic
in the extended zone scheme.  Moreover, new branches of surface wave appear
besides the continuum Rayleigh branch.  This richer band structure is crucial in
understanding features on resonant phonon emission.

\subsection{Theoretical surface dispersion relations}

Due to the simplicity of the model it is not difficult to evaluate the dispersion relations of its surface waves.
These are solutions of $\omega^2 {\bf u}_{\bf k} + \nabla\cdot\stackrel{\leftrightarrow}{\sigma} = 0$
with boundary condition 
$\stackrel{\leftrightarrow}{\sigma} \cdot {\bf n}=0$, where ${\bf n}$ is a unit 
vector normal to the surface.
Solutions are linear combination of bulk modes characterized by the same 
frequency $\omega$ and surface wave vector components, and different complex 
${\bf k}\cdot{\bf n}$ that give decay towards the interior of the 
sample.

Calculations of the surface modes are reported in appendix \ref{app:surf-modes}.
Results are shown in fig.~\ref{fig:s-disprel}.(a).

The figure shows quite a rich diagram which includes the Love branch 
and some high-frequency branches.  It is clear that all the branches 
could take part in the process of phonon emission.

%
%

\begin{figure}[b!]
\centerline{
\epsfxsize=8.5cm{\epsfbox{\dir 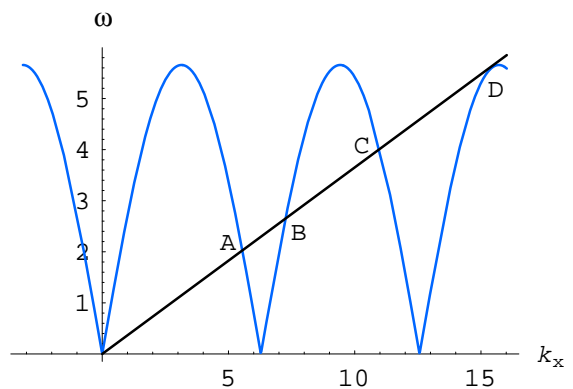}}}
\medskip
\caption{Schematic description of the basic matching phenomena.  A crack 
advancing at some speed $v$ matches wave phase velocity at points A, B, C and D, 
so that corresponding phonons can be emitted from the crack tip.  The figure shows 
the special case in which there is a \emph{resonant} match at D: the crack 
speed also matches the wave's group velocity which should lead to strong 
emission.}
\label{fig:surf-phenom} 
\end{figure}

\begin{figure*}
\centerline{
\epsfxsize=8.5cm{\epsfbox{\dir 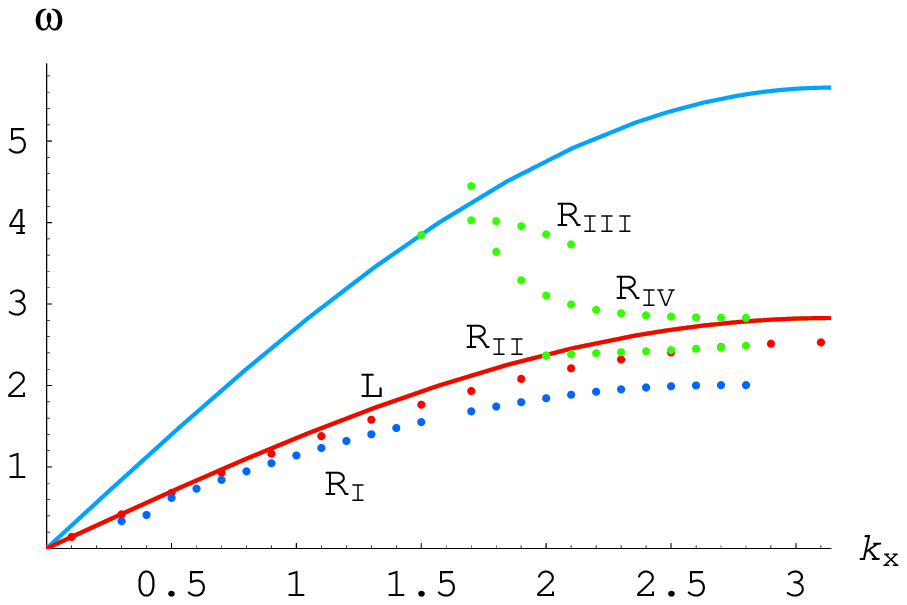}}\hspace{1cm}
\epsfxsize=9cm{\epsfbox{\dir 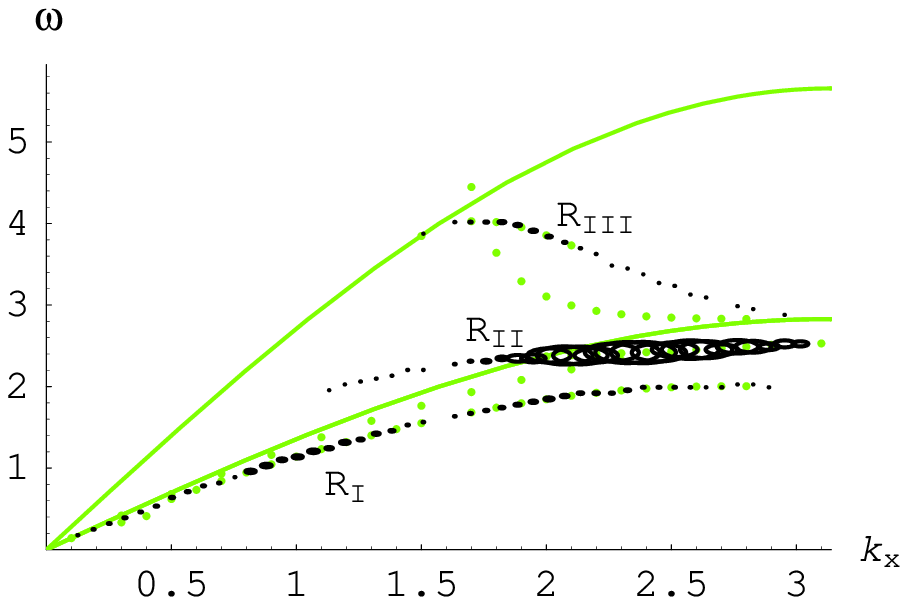}}}
\medskip
\caption{(a) Points represent theoretical dispersion relations for surface 
waves in the first Brillouin zone, for our model with a $(0, 1)$ surface normal
and $\nu=1/3$, with units as in figure \ref{fig:FCClattice-disp-rel}.  Lines show the 
dispersion relations for bulk modes.  All surface modes have $\textrm{Im}(k_y)>0$ 
to be damped for $y\rightarrow\infty$, and $k_z = 0$ corresponding to the 
two dimensional constraint in our simulations.  The Rayleigh and Love 
branches are labelled as R$_{\mbox{\scriptsize{I}}}$ and L.  The other high 
frequency branches are not interpretable in the continuum limit.
(b) Dispersion relations measured by excitation of a free surface, 
compared with the theoretical branches. More than one frequency can be excited 
for a given $k_x$.  The size of circles reflects the intensity of the signals.
Only three of the five branches are excited: the Love branch is not visible due
to its polarization, however it is not clear the reason of the lack of 
evidence of the R$_{\mbox{\scriptsize{IV}}}$ branch.}
\label{fig:s-disprel} 
\end{figure*}


\subsection{Direct measurement of surface waves}

The dispersion relations can be directly measured by exciting the top free 
surface of a simulated elastic sample.  Sites of the top face are displaced 
according to one specified wave vector $k_x$.  The surface is let 
free to oscillate and the frequencies of the corresponding excited modes 
are measured.  
Changing the $k_x$ vector at $k_z=0$, the entire first Brillouin zone can be 
spanned and a 
direct measurement of the dispersion of surface waves can be obtained.  Results 
are shown in figure \ref{fig:s-disprel}.(b).

It is evident that all the surface excitation is associated with the theoretical 
branches.  No emission corresponding to the Love branch is visible due to its 
polarization perpendicular to our two dimensional plane.
We have not identified the reason why no emission is observed corresponding 
to the R$_{\mbox{\scriptsize{IV}}}$ branch.

Our measurements do show continuation of the theoretical branches beyond their end 
points.  We have verified from calculations of appendix \ref{app:surf-modes} 
that these continuations correspond to frequencies with a small imaginary part.

%
%

\subsection{Emission from advancing cracks}

\begin{figure}[b!]
\centerline{
\epsfxsize=9cm{\epsfbox{\dir 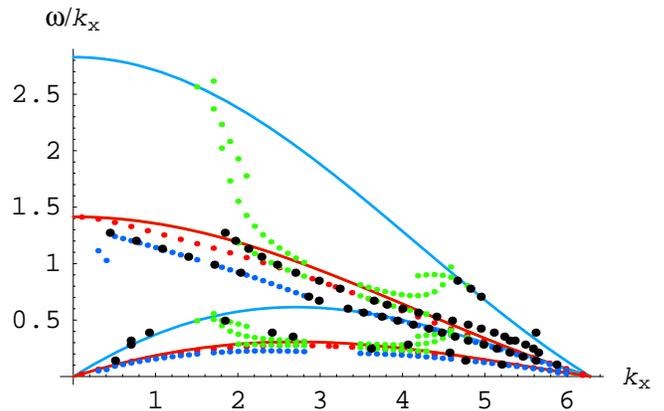}}}
\medskip
\caption{In this plot crack speeds correspond to horizontal lines.  Black 
dots 
correspond to the emission measured from the crack surface.  For simplicity only 
emission from the first and second Brillouin zone (folded back into the first 
zone) has been reported. All dots lay on the theoretical dispersion relations or 
on their extensions in the second Brillouin zone.}
\label{fig:crack-emiss} 
\end{figure}

Let us consider a snapshot at time $t$ of the crack surface.  Depending on the 
crack speed, some of the vibrational modes will be excited.  Through a Fourier 
transform of the surface profile the excited $k$-vector can be obtained.  
Assuming that eq.~(\ref{eq:cherenkov}) applies we can compare observed values 
of $k_x$ at crack speed $v$ with $k_x$ versus $\omega / k_x$ from the 
theoretical dispersion relations.  The one delicate point is that $k_x$ is 
indeterminate up to multiples of a reciprocal lattice vector $g$.  
Figure \ref{fig:crack-emiss} shows the agreement 
between measured emission and the theoretical curves, with assignment of 
multiples of $g$ to the measured $k_x$ as the only element of fitting.

%
%

\subsection{Resonances}

\begin{figure}[b!]
\centerline{
\epsfxsize=7.5cm{\epsfbox{\dir 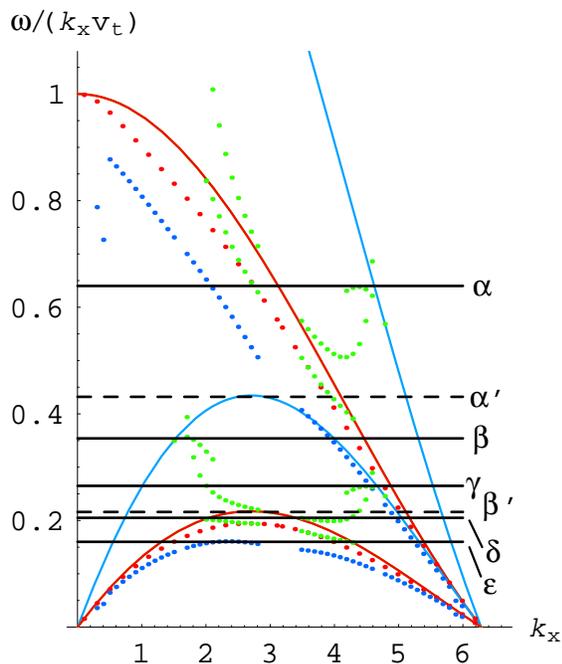}}}
\medskip
\caption{The theoretical phase velocity $\omega/k_x$ versus $k_x$, with horizontal 
lines showing where resonant phonon emission can be expected.  The lines show resonances 
due to the first and the second Brillouin zone only.  Dashed lines correspond to 
resonances due to bulk dispersion relations.}
\label{fig:theory-resonances} 
\end{figure}

Resonances appear as soon as eq.(\ref{eq:resonant-condition}) is fulfilled:
this matching of phase velocity and group velocity can lead to resonant emission 
as sketched in fig.~\ref{fig:surf-phenom}.  
Plotting the phase velocity $\omega/k_x$ versus $k_x$, the resonant 
condition is equivalent to a plateau in $\omega / k_x$.
Predicted resonances can therefore be easily read from our computed dispersion 
relations as shown in fig.~\ref{fig:theory-resonances}.
Due to the periodic band structure there is an infinite number 
of possible resonances corresponding to the infinite number of Brillouin zones.  
However we observe (not surprisingly) that signal 
corresponding to resonances in higher Brillouin zones is weak, so our discussion
in this paper focuses on the first two Brillouin zones only.

\begin{figure}
\centerline{
\epsfxsize=9cm{\epsfbox{\dir 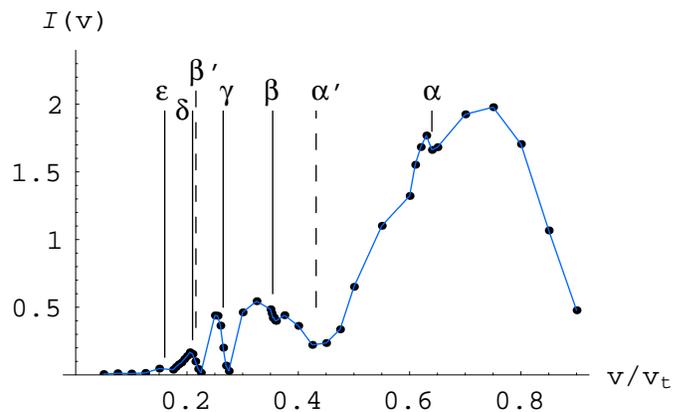}}}
\medskip
\caption{The total wave intensity integrated over the fracture surface, as a function 
of crack speed.  Peaks in the emission correspond to the resonances shown in figure 
\ref{fig:theory-resonances}, as indicated by vertical lines.  A systematic shift 
towards the lower crack speeds is 
visible and is related to the convexity of the dispersion relations (see text).
These results were obtained from simulations of samples of $800 \times 800$
tetrahedra.}
\label{fig:tot-power} 
\end{figure}

The presence of resonances can be verified by analysing the total 
intensity 
of emitted waves as a function of crack speed.  Figure \ref{fig:tot-power} shows the total
wave intensity integrated over just the crack surface, whilst the discussion of 
the full radiated wave power will be given in section \ref{sec:energy-release-rate}.

Corresponding to the speeds where resonances are expected there is an effective 
peak in the wave intensity.  These peaks are slightly shifted towards lower 
speeds, which can be interpreted in terms of the concavity 
of dispersion relations at the resonance.  When the crack advances at a 
speed slightly below the resonance, more modes can be excited, giving a maximum 
in intensity slightly shifted.  The abrupt drop of intensity at higher crack speeds 
(approaching the Rayleigh speed) appears to result from the fact that the 
penetration length of such waves is of order their wavelength, and this 
in turn becomes of the order of the sample size as the Rayleigh speed is approached.  
At high speeds, the total intensity of surface waves cannot be retrieved by 
analysing just the crack profile.


%
%

\section{ENERGY RELEASE RATE}
\label{sec:energy-release-rate}

The presence of resonances suggests that an effect on the crack dynamics 
has to be expected.  We will now show how the most commonly and relevant 
measured quantity of crack dynamics, namely the \emph{energy release rate}, 
is affected by this phenomenon.

In the simulations here reported, a planar crack is left to advance at 
constant speed from a short starting notch.  At early times the crack 
is almost equivalent to a crack advancing in an unbounded medium, since 
the sample boundaries are distant from the crack tip.
When the crack becomes large compared to the linear 
dimensions of the sample the crack starts to ``feel'' the presence of the 
boundaries and the correct description is that of a crack advancing in a strip.
In the continuum and steady limit the short crack regime is characterized by a 
linear increase of the energy release rate $G$ with time, whilst in the long 
crack regime the $G$ function is time-independent.  Both regimes 
are sensitive to the crack speed.

\subsubsection{Energy release rate in a discrete sample}

The discrete nature of matter is reflected in the dependence of the energy 
release rate $G$ on crack speed. We can describe the macroscopic energy
release rate per unit distance of crack advance $G_M$ as the sum of 
two contributions:
\begin{equation}
\label{eq:ansatz-G}
G_M(v,t) = G_{\textrm{\small{br}}}(v,t) + G_{\textrm{\small{ph}}}(v,t).
\end{equation}

\noindent
$G_M(v,t)$ is the solution of the continuum limit which governs the 
macroscopic delivery of energy towards the crack tip, for which we have
theoretical expressions available.  
$G_{\textrm{\small{br}}}(v,t)$ and $G_{\textrm{\small{ph}}}(v,t)$ are 
respectively the breakage energy release rate and the phonon energy release rate. 
Our strategy below is to directly measure the breakage energy release rate 
from the potential energy lost when tetrahedra are broken.

The macroscopic energy release rate $G_M(v,t)$ is largely determined by
the macroscopic conditions and the length of the crack.  
In the long crack limit in the case of a strip of height $2l$ and 
fixed displacement $\delta$ at each boundary, the macroscopic energy 
release rate corresponds to the amount of elastic energy stored 
far ahead of the crack tip\cite{F-90,N-72}.  
From eq.(\ref{eq:stress-strain}) the stress field ahead of the 
crack tip has
\begin{equation}
\label{eq:sigma-stripe}
\sigma_{yy} = (\lambda+2\mu)\epsilon
\end{equation}

\noindent
where $\epsilon = \delta/l$ is the imposed strain.  Hence the
time independent macroscopic energy release rate is:
\begin{equation}
\label{eq:G_M_infty}
G^{\infty}_M = (\lambda + 2\mu) \epsilon^2 l
\end{equation}

\noindent
independent of the crack speed $v$.

\begin{figure}[t!]
\centerline{
\epsfxsize=8cm{\epsfbox{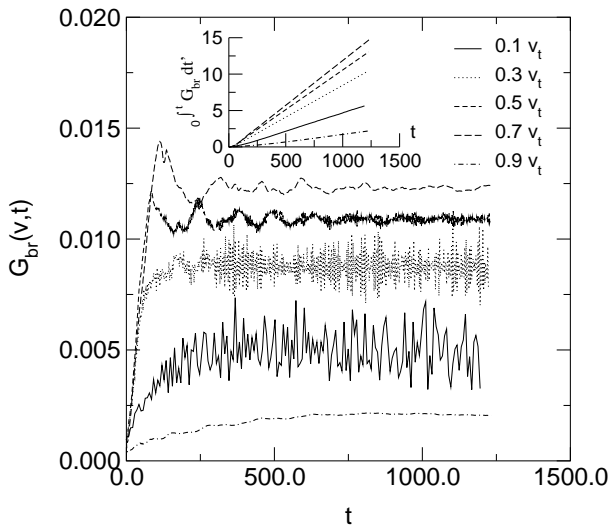}}}
\medskip
\caption{The measured energy delivered into bond breaking per unit length of crack,
as a function of time for different crack speeds. 
From these data we measured the early time slopes $G_{\textrm{\small{br}}}'(v,0)$
corresponding to the short crack regime and (from the slope of the time integrated 
plots inset) the long crack plateau values $G_{\textrm{\small{br}}}(v)$.
The (non-trivial) sequence of the 
curves is more readily appreciated from fig. \ref{fig:G-resonances}.
}
\label{fig:G-measure-nodamp} 
\end{figure}

The translation of the Griffith criterion in the discrete case
is that the crack will advance as soon as the energy stored in the
tetrahedron to break, $G_{\textrm{\small{br}}}$,
is greater than a threshold value connected with the toughness
of the material.  
The behaviour of $G_{\textrm{\small{br}}}$ with the
crack speed is therefore crucial for understanding the crack dynamics.

The breakage energy release rate clearly depends 
on the external macroscopic conditions as well as the discreteness
of the model.  However it can be expressed as:
\begin{equation}
\label{eq:def-efficiency}
G_{\textrm{\small{br}}}(v,t)=
   E(v)\,G_M(v,t) 
\end{equation}

\begin{figure}[t!]
\centerline{
\epsfxsize=8cm{\epsfbox{\dir 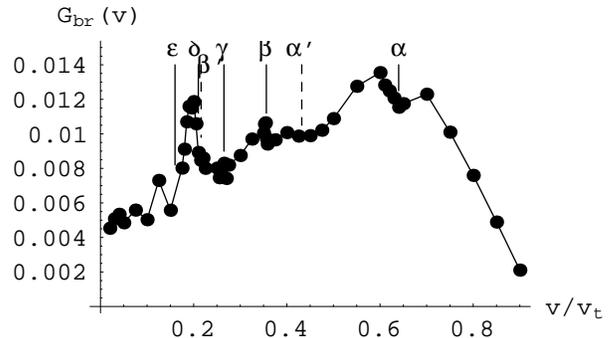}}}
\medskip
\protect \caption{Dependence of breakage energy release rate on crack speed in
the long crack regime.  Crack propagation with a steady speed should only be 
stable where this function is decreasing.  Vertical lines show the theoretical position 
of resonant surface wave emission: the two dashed lines correspond to resonant bulk wave 
emission.  The figure equivalently shows the behaviour of the efficiency $E(v)$ with 
the crack speed, as for long cracks and fixed grip conditions the two are strictly 
proportional.}
\label{fig:G-resonances} 
\end{figure}

\noindent
where we have introduced the \emph{efficiency} $E(v)$.  In the 
long crack limit for the fixed grip set-up used in our simulations
this gives $G_{\textrm{\small{br}}}(v,t)=E(v)\,G^{\infty}_M$
where the efficiency is the sole source of velocity dependence.
The effect of local discreteness thus separates from the 
effect of macroscopic external conditions: all the dependence
on the crack speed is hidden in the efficiency function $E(v)$
which is local to the crack tip region and independent of the
macroscopic regime as will be shown below.
The meaning of the efficiency $E(v)$ is as follows: when $E(v)$
is close to zero, the energy delivered to the crack tip is 
mostly spent in phonon emission so that the mechanism is not
sufficient for the crack to advance.  When the $E(v)$ is close
to one, all the energy delivered is used to break tetrahedra, and
the crack can advance promptly.
We will see that the dependence on $v$ is crucial in the determination
of bands of permitted crack speeds, not described in the continuum 
elastic theory.  The correct derivation of these bands will be 
given below.

\subsubsection{Measuring the energy release rate}
\label{sec:results}

The breakage energy release rate $G_{\textrm{\small{br}}}(v,t)$ in our 
simulations corresponds, for a given crack moving at speed $v$, to 
the value of the elastic energy that disappears from the system with each 
broken tetrahedron at time $t$.
Figure \ref{fig:G-measure-nodamp} shows measurements from a set of simulations 
involving samples of of height $120$ tetrahedra 
and up to $3500$ tetrahedra long for increasing crack speeds. 
We observe $G_{\textrm{\small{br}}}(v,t)$ to grow linearly with time initially:
corresponding measured slopes $G_{\textrm{\small{br}}}'(v,0)$ are discussed
below.
The long crack regime exhibits fluctuations due to waves reflected from the 
sample boundaries, but the average magnitude $G_{\textrm{\small{br}}}(v)$ 
of the breakage energy release rate can be clearly retrieved.

The $G_{\textrm{\small{br}}}(v)$ in the long crack 
limit is shown as a function of crack speed in figure \ref{fig:G-resonances}.  
Vertical lines indicate the speeds at which resonances 
are expected to be seen according to figure \ref{fig:surf-phenom}.
The two lines labelled as 
$\alpha'$ and $\beta'$ correspond to resonances due to the dispersion relations 
of \emph{bulk waves}.
Corresponding to each line there is a clear decrease of the energy release rate, 
meaning that more energy is emitted as radiation.

The crucial feature of figure \ref{fig:G-resonances} is that it reveals how the 
energy available for breaking bonds responds to the speed of the crack, for a given
macroscopic energy release rate.
This same figure can be read in reverse: given some threshold value for
$G_{\textrm{\small{br}}}(v)$ corresponding to a given fracture toughness of 
the material, the possible crack speeds are obtained from the graph.

It can be further argued on grounds of stability that only the speed ranges
where $G_{\textrm{\small{br}}}(v)$ decreases with $v$ are allowed as steady 
crack speeds.  
In the counter case where $G_{\textrm{\small{br}}}(v)$ increases with $v$, a 
prospective overshoot in $v$ leads to excessive bond breakage energy and hence 
acceleration, and an undershoot to insufficient breakage energy and the crack 
must slow further.
Thus stable steady crack propagation is confined to narrow intervals associated 
with resonances and the high speed regime.

It is worth pointing out that most of the features shown in figure 
\ref{fig:G-resonances} are due to high frequency branches either in 
the first or second Brillouin Zone.  The first resonance due to the Rayleigh 
branch (apart from the resonance at the Rayleigh speed) is the marginally visible 
$\epsilon$ resonance.  This shows how important it is to include the 
full complexity of the band structure within the analysis.

The data in fig.~\ref{fig:G-resonances} are at constant macroscopic energy release 
rate $G_M^{\infty}$, so due to (\ref{eq:def-efficiency}) they also show the 
behaviour of the efficiency $E(v)$ with the crack speed.
However, if the efficiency $E(v)$ is governed only by local phenomena, then 
its behaviour with the crack speed should be independent of the crack propagating 
in steady state or transient regime, provided the crack speed $v$ is fixed.  
From the knowledge of $E(v)$ then we could
find out the dependence of $G_{\textrm{\small{br}}}(v,t)$ on the crack speed
for any macroscopic set up, provided we know the macroscopic continuum solution.

To verify the independence of $E(v)$ on the dynamic regime, we also analyzed the 
short crack limit.  In this case, the crack can be seen as advancing in
an unbounded medium as the sample boundaries are distant from the crack tip.
For a planar type-I crack propagating at steady speed $v$ in an unbounded 
continuum medium the macroscopic transfer of energy to the crack tip (per unit distance 
advanced) is given by:
\begin{equation}
\label{eq:G-unbounded}
G^0_M(v,t) = \displaystyle\frac{v^2 \alpha_l}{2 c_t^2 \mu D} K^2_I(v,t)
.
\end{equation}

\noindent
Here $\alpha_{t,l} = \sqrt{1 - v^2/v^2_{t,l}}$ and 
$D=4\alpha_t\alpha_l - (1+\alpha^2_t)^2$.  
The constants $v_l$ and $v_t$ represent the longitudinal and 
transverse sound speeds, and $K_I(v,t)$ is the \emph{stress intensity factor} at 
time $t$ for the given crack speed.

A functional form for $K_I(v,t)$ is available for the Broberg 
problem\cite{F-90,B-60,B-99} of a crack expanding from zero initial length 
in a uniform tension field and infinite medium.  In particular it can 
be written in the form:
\begin{displaymath}
K_I(v,t) = \Sigma(v) \sigma_{\infty}\sqrt{\pi vt}
\end{displaymath}

\noindent
where $\sigma_{\infty}$ is the traction applied on the crack faces.  
The energy release rate increases linearly with time.
As figure \ref{fig:G-measure-nodamp} shows, we obtain the same behaviour 
for $G_{\textrm{\small{br}}}(v,t)$ in our simulations.
In the case of symmetric growth $\Sigma(v)$ has the form:

\begin{displaymath}
\Sigma(v)=-\frac{I(b/q)R(q)}{b^2q \sqrt{q^2-a^2}}
\end{displaymath}

\noindent
where $R(q) = (b^2-2q^2)^2+4q^2 \sqrt{a^2-q^2}\sqrt{b^2-q^2}$ and 
\begin{displaymath}
I^{-1}(b/q) = \frac{q}{b^2} \int_0^{\infty}\!\! \frac{R( i \eta )}{(q^2+\eta^2)^{3/2} \sqrt{a^2+\eta^2}}\ d\eta
\end{displaymath}

\noindent
with $a=v_l^{-1}$, $b=v_t^{-1}$, $q=v^{-1}$.  We thus have a closed form for
the early time slope $G_M'(v,0)$.

Due to these results, to prove that $E(v)$ is independent of the dynamic 
conditions we just have to check that
\begin{equation}
E(v,0) = \frac{G_{\textrm{\small{br}}}'(v,0)}{G_M'(v,0)}
\end{equation}

\noindent
matches $E(v)$ already measured from the long crack limit.
Simulations of symmetric 
crack growth can be performed simply by fixing the longitudinal 
displacement at the left and right boundaries:  
though this would correspond to having a periodic system of symmetric 
growing cracks, the interaction between cracks is weak due to 
the strip geometry.
$G_{\textrm{\small{br}}}'(v,0)$ then corresponds to the measured slopes 
of the early regime, and $G_M'(v,0)$ is the time derivative of the 
macroscopic energy release rate (\ref{eq:G-unbounded}).

Results of the comparison are reported in table \ref{tab:tab1}.
Because the product $vt$ corresponds to the crack length, the case $v=0$ was 
reconstructed by analysing several simulations of static cracks with 
different crack lengths and retrieving the resulting overall slope.
For low and high crack speeds, problems arise when measuring the short crack regime:
for low crack speeds data is noisy and there are few points to average; 
for high crack speeds the long crack limit is achieved soon after the 
crack starts moving, so that the short crack regime is properly defined 
only when the sample height is very large with respect to the size of 
a tetrahedron.  This is still not the case even when dealing with our 
$120$ tetrahedra wide sample.  Better results for the short crack regime 
were obtained from simulations of $800\times 800$ tetrahedra as the third 
column of the table shows.  

\begin{table*}[t!]
\begin{ruledtabular}
\begin{tabular}{c c c c c}
  & $2l = 120$ & $2l = 800$ & $2l = 120$ & \\
  & & & & \\
$v/v_t$ & 
  $E(v,0)= \displaystyle\frac{G_{\textrm{\scriptsize{br}}}'(v,0)}{G_M'(v,0)}$ & 
  $E(v,0)= \displaystyle\frac{G_{\textrm{\scriptsize{br}}}'(v,0)}{G_M'(v,0)}$ & 
  $E(v)  = \displaystyle\frac{G^{\infty}_{\textrm{\scriptsize{br}}}(v)}{G^{\infty}_M(v)}$  &
  $E(v,0)/E(v)$ \\
  & & & & \\
\hline
$0.0$ &  $0.257 \pm 0.006$  & -                 & $0.2591 \pm 0.0008$  & - \\
\hline
$0.1$ &  $0.19  \pm 0.04 $  & $0.22  \pm 0.04 $ & $0.2408 \pm 0.0004 $  &  $0.9   \pm 0.2  $  \\
$0.2$ &  $0.52  \pm 0.03 $  & $0.52  \pm 0.01 $ & $0.528  \pm 0.003  $  &  $0.98  \pm 0.02 $  \\
$0.3$ &  $0.40  \pm 0.02 $  & $0.390 \pm 0.006$ & $0.379  \pm 0.001  $  &  $1.03  \pm 0.02$   \\
$0.4$ &  $0.44  \pm 0.02 $  & $0.432 \pm 0.004$ & $0.4315 \pm 0.0002 $  &  $1.00  \pm 0.01$   \\
$0.5$ &  $0.48  \pm 0.02 $  & $0.462 \pm 0.004$ & $0.4636 \pm 0.0001 $  &  $0.997 \pm 0.009$  \\
$0.6$ &  $0.59  \pm 0.02 $  & $0.576 \pm 0.006$ & $0.5767 \pm 0.0004 $  &  $1.00  \pm 0.01$   \\
$0.7$ &  $0.55  \pm 0.04 $  & $0.514 \pm 0.006$ & $0.5201 \pm 0.0001 $  &  $0.99  \pm 0.01$   \\
$0.8$ &  $0.2   \pm 0.1  $  & $0.33  \pm 0.02 $ & $0.3202 \pm 0.0001 $  &  $1.03  \pm 0.06$   \\
$0.9$ &  -                  & $0.079 \pm 0.007$ & $0.0890 \pm 0.0005 $  &  $0.89  \pm 0.08$
\end{tabular}
\end{ruledtabular}
\medskip
\caption{The efficiency of energy delivery into bond breaking, 
compared between the different dynamical regimes of short and long cracks, 
as a function of crack speed.  The second and third columns show the efficiency
computed from early time slopes of the energy into bond breakage compared to 
that macroscopically delivered, for two different sample heights.
Results for $v=0$ were obtained from a set of simulations of static 
fracture with different notch lengths. Results for low and high crack 
speeds are influenced by problems in the measurements in the short 
crack limit: this effect is reduced increasing the size of the sample 
as shown in the third column.  The final column shows the ratio of these 
short crack regime efficiencies to those computed from the long crack regime
shown in the fourth column:  these results clearly confirm that the
efficiency is insensitive to dynamical regime, consistent with our
assertion that it is a property local to the crack tip.}
\label{tab:tab1}
\end{table*}

These results show that the efficiency $E(v)$ is independent of the particular
dynamic regime.  
The absolute values are also of interest.  The $v=0$ limit shows that when the
tetrahedra at the tip of a static crack are broken, about $75\,\%$ of the strain 
energy released comes from relaxation in other tetrahedra which is radiated in waves.
The maximum efficiency (about $60\,\%$) occurs at $v \simeq 0.6\,v_t$.
This happens to be just below the Yoffe speed $\sim 0.63\,v_t \simeq 2/3 v_R$,
but as our measurements exclude the possibility of crack branching we presume this 
to be a coincidence.

%
%

\section{CONCLUSIONS}

We have presented a new finite element model for linear elastic fracture
mechanics, which has proved surprisingly and revealingly rich in its behaviour 
even in the two dimensional case.  The model was designed to enable fast numerical 
simulations of large systems particularly with three dimensions in mind, but the 
main strengths which we have exploited in this paper are direct control of the 
local physics which it offers, combined with bulk and surface
dispersion relations amenable to simple theoretical computation.
Our results suggest that inclusion of known phonon dispersion can be crucial to 
understanding the speed of fracture propagation.

Our results relate to ideally brittle materials, in that we have included no 
significant mechanism of local dissipation and most particularly no plastic 
deformation mechanism.  Linear damping is readily included, and indeed could be 
exploited to mitigate the effects of waves reflected back from the sample edges.
Spontaneous crack roughness and branching will be addressed in a following paper.

The crucial mechanism which our results incorporate beyond continuum fracture 
mechanics is the radiation of phonons from the crack tip. This we show leads to 
a significant speed dependence in the fraction of macroscopic strain energy available
as work to create new surface.  For a static crack in our model, this efficiency is 
only $25\,\%$, thus modifying the most na\"{\i}ve Griffith criterion for crack 
propagation by a factor of $4$.  The general rise in bond breaking efficiency with 
speed towards a global maximum for typical fast crack speeds rules out steady crack
propagation at most lower speeds, and remains to be understood more quantitatively.

We have been able to interpret fine structure in the bond breaking efficiency 
associated with resonant conditions for surface wave emission.  This leads to islands
of stable crack velocity, which can only arise at lower speeds due to dispersion and
hence due to the discrete properties of matter.

A related explanation of the presence of sets of forbidden crack speeds 
has already been given by 
M.~Marder \emph{et al.}\cite{MG-95,HM-99,FM-99} by using a mechanistic 
description of bond breakage in terms of most stretched ones.  According to 
this approach cracks cannot advance at speeds below some threshold, since due 
to lattice oscillations bonds would break before the expected time compatible 
with the crack speed.  A velocity gap then appears.  The two approaches seem 
to be very close indeed, since simulations from Murder \emph{et al.} are based 
on periodic lattices which should show characteristic dispersion relations.  
The oscillations between neighbouring sites which stretch the bonds beyond the 
critical length correspond to wave vectors at the border of the Brillouin zone 
and ideally should be connected to the presence of resonances towards the zone 
boundary.
Our description however appears to be more general as it expresses the same 
phenomenon in terms of the energy release rate and phonon band structure.  
The existence of velocity gaps is shown on the basis of energetic arguments.  
This has the advantage of not being built up on the particular rule for 
breaking bonds, but relates the existence of velocity gaps and constant speed 
advance to a more general description of the properties of the material.  
Furthermore, velocity gaps can be read directly from the dependence of the 
strain energy release rate on the crack speed, a relationship which we might
eventually hope to deduce or calculate for real materials.

\begin{acknowledgments}
AP would like to thank A.Petri for its observations and the useful discussions.
We aknowledge support of EU contract n. ERBFMRXCT980183
\end{acknowledgments}


\appendix

%
%

\section{Dispersion relations for bulk waves}
\label{app:bulk-waves}

We need to find the eigenvalues for eq.~(\ref{eq:motion-model}).
By inspection ${\bf u}_{\bf k} = \nabla \wedge \nabla^*$
is an eigenvector with eigenvalue
\begin{equation}
\label{eq:bulk-mode1}
\textrm{A}_{\bf k} \equiv 
\frac{\mu\,\Omega'}{m}[\nabla\cdot\nabla^*+\nabla^*\cdot\nabla]
.
\end{equation}

\noindent
The remaining two eigenvectors are given by a linear combination of 
$\nabla$ and $\nabla^*$.
We then obtain:
\begin{eqnarray}
\label{eq:tetmod-dis-first}
\lefteqn{\omega^2(b\nabla+c\nabla^*) = \textrm{A}_{\bf k}(b\nabla+c\nabla^*) 
+{} } \nonumber\\
& & \quad + \frac{(\lambda+\mu)\,\Omega'}{m}[\nabla\nabla^*+\nabla^*\nabla]\cdot(b\nabla+c\nabla^*)
.
\end{eqnarray}

\noindent
We can now multiply eq.~(\ref{eq:tetmod-dis-first}) by $\nabla$ to the left and 
by $\nabla^*$ to the right, so as to obtain the following coupled equations:
\begin{displaymath}
\left\{
\begin{array}{lll}
(\epsilon - 1) [b\eta + c ] & = & \beta [2b\eta + c(1+|\eta|^2)]\\
  & & \\
(\epsilon - 1) [b + c\eta^* ] & = & \beta [b(1+|\eta|^2) + 2c\eta^*]
\end{array}
\right.
\end{displaymath}

\noindent
with: $\epsilon \equiv \omega^2 / \textrm{A}_{\bf k}$,
$\beta \equiv (\lambda+\mu)\,\Omega'\,\nabla\cdot\nabla^* / (m\,\textrm{A}_{\bf k})$
and $\eta  \equiv (\nabla\cdot\nabla) / (\nabla\cdot\nabla^*)$.
Solving for eigenvalues leads to:
\begin{displaymath}
\epsilon = 1 + \beta(1\pm|\eta|) \qquad \textrm{with} \quad |\eta| \neq \mp 1
\end{displaymath}

\noindent
and substituting back into the equation we obtain:
\begin{equation}
\label{eq:tetmod-eigenvalues}
\omega^2 = \textrm{A}_{\bf k} + \frac{(\lambda+\mu)\,\Omega'}{m} \Big[(\nabla\cdot\nabla^*) \pm
           |\nabla\cdot\nabla| \Big]
\end{equation}

\noindent
which together with (\ref{eq:bulk-mode1}) completes the dispersion relations.

Having obtained these eigenvalues, it is not difficult to find that 
the corresponding eigenvectors have the form:
\begin{equation}
{\bf u} = \nabla \pm \displaystyle\frac{\nabla\cdot\nabla}{|\nabla\cdot\nabla|}
          \nabla^*
.
\end{equation}

%
%

\section{Adding dissipation}
\label{app:damp}

There are at least two pratical reasons to add dissipation in the dynamics.
First, the starting point of any simulation should be a sample in equilibrium:
this can be easily obtained by relaxing the lattice to a configuration of 
minimum energy.
The second reason is the possibility to damp out waves that would otherwise propagate 
and reflect back from the sample borders. 

The physical way to introduce dissipation in the model for linear elasticity is 
the introduction of a viscosity term in (\ref{eq:rho-sigma}).
However this gives wide dispersion in the damping rates of different 
modes making it inefficient at relaxing solutions or reducing 
boundary reflections.

\begin{figure}
\centerline{
\epsfxsize=7cm{\epsfbox{\dir 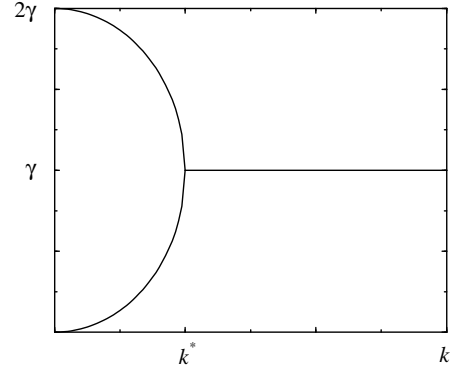}}}
\medskip
\caption{Damping rate for different values of $\gamma$.  Note that 
for $k>k^*$ all modes are damped at the same rate $q=\gamma$.}
\label{fig:disp-rel} 
\end{figure}

There is a less physical way to obtain dissipation that has 
the advantage of being less 
wave vector dependent and simpler to implement.  Starting from 
eq.~(\ref{eq:motion-model}) a dissipative term is added so that:
\begin{equation}
\label{eq:diss}
m \ddot{\bf u} + 2m\gamma\dot{\bf u} + \textrm{G}({\bf k}){\bf u} = 0
\end{equation}

\noindent where:
\begin{eqnarray*}
\lefteqn{\textrm{G}({\bf k}){\bf u} = \Omega'\Bigg\{ -\mu \Big[\nabla \cdot \nabla^* + 
         \nabla^* \cdot \nabla \Big] {\bf u} +}\hspace{1cm}\nonumber\\ 
 & \qquad & - (\lambda + \mu) \Big[\nabla \nabla^* +
                    \nabla^* \nabla \Big] \cdot {\bf u} \Bigg\}
.
\end{eqnarray*}

\noindent
This approach is clearly unphysical, but useful to reach the configuration 
of minimal energy or to damp out unwanted reflected waves.

Solutions of the form ${\bf u}(t) = e^{-qt} {\bf u}_{\bf k}$ can be found by 
substitution in (\ref{eq:diss}), giving
$q^2 - 2 \gamma q + \textrm{G(} {\bf k} \textrm{)} / m = 0$,
and leading to

\mbox{}

\begin{displaymath}
\begin{array}{rcl}
\textrm{Re} \{ q \} & = & \textrm{Re} \Bigg\{ \gamma \pm \sqrt{ \gamma^2 - 
            \Bigg[\frac{\textrm{G(} {\bf k} \textrm{)} }{m} \Bigg] }
            \Bigg\} = \\
   &   & \\
   &   & \hspace{-1cm} =
\left\{ 
\begin{array}{l c c c c}
\gamma &   & \frac{\textrm{G(} {\bf k} \textrm{)} }{m} &\geq &\gamma^2\\
       &   &          &      & \\
\left\{

\begin{array}{l l}
2 \gamma - \frac{1}{2 \gamma} 
         \Bigg[\frac{\textrm{G(} {\bf k} \textrm{)} }{m} \Bigg]
  & 
\begin{array}{l}
\textrm{\small{fast}}\\
\textrm{\small{modes}}
\end{array}\\
  & \\
\frac{1}{2 \gamma} 
       \Bigg[\frac{\textrm{G(} {\bf k} \textrm{)} }{m} \Bigg]  & 
\begin{array}{l}
\textrm{\small{slow}}\\
\textrm{\small{modes.}}
\end{array}

\end{array}

\right.
       &   & \frac{\textrm{G(} {\bf k} \textrm{)} }{m} & \ll & \gamma^2
\end{array}
\right.
\end{array}
\end{displaymath}

\noindent
In fig.~\ref{fig:disp-rel} the damping rate $\textrm{Re}\{q\}$ is plotted as a function 
of $|{\bf k}|$.
Curves depend on the value of the dissipative parameter $\gamma$: in particular, 
the value $k^*(\gamma)$ separates two regimes.
For $k > k^*$ all the modes are damped with a damping constant $q=\gamma$, 
whereas 
for $k < k^*$ fast and slow modes are damped with different magnitudes.
The optimal value for $\gamma$ is then set by maximising the damping rate of 
the slowest mode, leading us to set 
$k^* \equiv k_{\textrm{\small min}} = \frac{\pi}{L}$, 
where $L$ is the size of the lattice in fcc unit cells.
The choice is also interesting in that it sets all modes to have the same
damping rate
\begin{equation}
\label{eq:dissipation-gammachoice}
\gamma = c_t\frac{\pi}{L}
.
\end{equation}

\noindent
The time needed to reduce the potential energy of a factor $\epsilon$ towards 
its minimum is then given by:
\begin{equation}
\label{eq:dissipation-Delta_t}
\Delta t = - \frac{h}{2 \pi c_t} \ln \epsilon
.
\end{equation}

During the dynamics, reflected waves travel a distance of at least $L$ fcc
unit cells in a 
time interval of $\Delta t = L/c_t$, hence when reflected waves reach the 
crack tip, using the optimal value for $\gamma$ their intensity is reduced at least 
by a factor $e^{2 \gamma \Delta t} = e^{2 \pi} \sim 500$ for transverse waves, 
and $e^{2\pi \frac{c_t}{c_l}} \sim 25$ for longitudinal waves, using 
values as in figure \ref{fig:FCClattice-disp-rel}.  Note however that 
for the results presented in this paper we turned all damping off after 
initial equilibration.

\section{Dispersion relations for surface waves}
\label{app:surf-modes}

Consider a solid occupying the portion of space defined by $y<0$ 
with the surface $y=0$ as its only boundary.
In the continuum case, surface modes are solutions of $\omega^2 {\bf u}_{\bf k} +
\nabla \cdot 
\stackrel{\leftrightarrow}{\sigma}=0$ constrained by the boundary condition 
$\stackrel{\leftrightarrow}{\sigma}\cdot{\bf n} = 0$ where ${\bf n}$ is the 
normal to the surface.
They consist of linear combinations of bulk waves with different complex $k_y$ 
(that accounts for their damping towards the interior of the sample) obeying the 
bulk waves dispersion relations and common $\omega$, $k_x$ and $k_z$.

We can apply the same principles to our discrete model considering that the layer of sites corresponding to the $y=0$ plane has no force acting on it 
from above.  It follows therefore:
\begin{equation}
\label{eq:boundary-cond}
\nabla_1^+ \cdot \stackrel{\leftrightarrow}{\sigma}_2 +
\nabla_2^+ \cdot \stackrel{\leftrightarrow}{\sigma}_1 = 0
\end{equation}

\noindent
where:
\begin{displaymath}
\left\{
\begin{array}{lll}
\nabla_1^+ & = & \displaystyle\sum_{\{v | r_{v,y}>0\}} {\bf r}_v 
                 \exp^{i {\bf q}\cdot {\bf r}_v}\\
           &   & \\
\nabla_1^- & = & \displaystyle\sum_{\{v | r_{v,y}<0\}} {\bf r}_v 
                 \exp^{i {\bf q}\cdot {\bf r}_v}\\
           &   & \\
\nabla_1   & = & \nabla_1^+ + \nabla_1^-
.
\end{array}
\right.
\end{displaymath}

\noindent
Hence:
\begin{displaymath}
\nabla^+({\bf k}) \equiv \nabla_1^+({\bf k}) = - \nabla_2^-(-{\bf k})
.
\end{displaymath}

Using eq.~(\ref{eq:sigma-1}), the boundary condition (\ref{eq:boundary-cond}) 
becomes:
\begin{displaymath}
\begin{array}{lll}
\lefteqn{\textrm{B}({\bf k}){\bf u}=\lambda (\nabla_2^+ \cdot \nabla_1 +
        \nabla_1^+ \cdot \nabla_2) {\bf u} +} \\
 & & \\
 & & \lambda (\nabla_1 \nabla_2^+ + \nabla_2 \nabla_1^+) \cdot {\bf u} +
     \mu (\nabla_2^+ \nabla_1 + \nabla_1^+ \nabla_2) \cdot {\bf u} = 0
.
\end{array}
\end{displaymath}

We can solve this equation by expressing ${\bf u}$ as a linear combination of 
bulk eigenvectors ${\bf u} = a{\bf e}_1 + b{\bf e}_2 +c {\bf e}_3$ with common
$\omega$, $k_x$ and $k_z$.  For simplicity we can reduce
to the case\footnote{We are interested in particular to the $<$100$>$ direction, 
however the more general case can be obtained by following the same steps.
}
$k_z = 0$.

\noindent
Bulk dispersion relations for $k_z = 0$ have therefore the form 
$\omega=f_{\alpha}(k_x,k_y)$.
These can be inverted such that $k_y = f_{\alpha}^{-1}(\omega,k_x)$.  Hence,
surface modes are solutions of the following equation:
\begin{displaymath}
\det\left(
\begin{array}{c}
\displaystyle\Big(\textrm{B}({\bf k}) {\bf e}_1\Big)_{k_y\rightarrow
              f_1^{-1}(\omega,k_x)}\\
              \\
\displaystyle\Big(\textrm{B}({\bf k}) {\bf e}_2\Big)_{k_y\rightarrow
              f_2^{-1}(\omega,k_x)} \\
              \\
\displaystyle\Big(\textrm{B}({\bf k}) {\bf e}_3\Big)_{k_y\rightarrow
              f_3^{-1}(\omega,k_x)}
\end{array}
\right) = 0
\end{displaymath}

\noindent
provided $\textrm{Im}\{f_1^{-1}(\omega,k_x),f_2^{-1}(\omega,k_x), f_3^{-1}(\omega,k_x) 
\} 
> 0$.


\end{document}